\documentstyle[12pt,epsfig,epsf]{article}
\def\ga{\mathrel{\raise.3ex\hbox{$>$\kern-.75em\lower1ex\hbox{$\sim$}}}}
\def\la{\mathrel{\raise.3ex\hbox{$<$\kern-.75em\lower1ex\hbox{$\sim$}}}}
\begin{document}

\newcommand{\nn}{\noindent}
\renewcommand{\thepage}{}
\renewcommand{\thefootnote}{\fnsymbol{footnote}}
\begin{center}
{\Large{ {\bf Single and pair production of MSSM Higgs bosons 
as a probe of scalar-pseudoscalar mixing at $e^+e^-$ colliders\footnote{Talk given at the 10th International Conference on Supersymmetry and Unification of Fundamental Interactions, DESY, Hamburg, Germany, June 17--23, 2002.}}}}

\vspace{.4cm}
{\large Andrew Akeroyd$^1$ and Abdesslam Arhrib$^{2,3}$\footnote{A. Arhrib is supported by Alexander von Humboldt Foundation.} }

\vspace{.4cm}
1: KIAS, 207-43 Cheongryangri--dong, 
Dongdaemun--gu, Seoul 130--012, Korea
\\
2: Max-Planck Institut f\"ur Physics, F\"ohringer Ring 6, 
80805 M\"unchen, Germany 

3: LPHEA, Physics Department, Faculty of Science, P.O.Box 2390, 
Marrakesh, Morocco.  
\end{center}

\begin{abstract}  
We study the associated production of the 
$A^0$ neutral CP--odd Higgs boson with a neutral gauge boson $Z$
as well as single production of $A^0$  via 
$e^+e^- \to \nu_e {\bar\nu}_e A^0$ at the one loop level 
in the Minimal Supersymmetric Standard Model (MSSM).  
We show that the MSSM cross--section 
may be enhanced by light SUSY particles. 
Then we study the production processes $e^+e^-\to 
H^0_iZ$,  $H^0_i\nu_e\overline \nu_e$ and $H^0_iH^0_j$
in the context of the MSSM with scalar-pseudoscalar mixing. 
In a given channel we show that the cross--section for all $i$
($=1,2,3$) can be above 0.1 fb provided $M_{H_{2,3}}\la 300$ GeV. 
This should be detectable at a Next Linear
Collider and would provide evidence for 
scalar--pseudoscalar mixing.
\end{abstract}

{\bf 1.} Recently, the phenomenology of the MSSM with complex SUSY parameters
has received growing attention \cite{MSSMCP2}.
Such phases give new sources of CP violation 
which may provide: electroweak baryogenesis scenarios  
and CP violating phenomena in K and B decays \cite{baek}.
It has been shown that by assuming universality of the 
gaugino masses at a high energy scale, the effects of complex soft SUSY 
parameters in the MSSM can be parametrized 
by two independent CP phases: the phase of the 
Higgsino mass term $\mu$ (Arg($\mu$)) and the phase 
of the trilinear scalar coupling parameters $A=A_f$ (Arg($A_f$)) 
of the sfermions $\widetilde{f}$. 
The presence of large SUSY phases can give contributions to the 
electric dipole moments of the electron and neutron (EDM) 
which exceed the experimental upper bounds. In a variety of SUSY 
models such phases turn out to be severely suppressed 
by such constraints i.e. ${\rm Arg}(\mu) < {\cal }(10^{-2})$ for a SUSY
mass scale of the order of few hundred GeV \cite{nath}.

However, the possibility of having large CP 
violating phases can still be 
consistent with experimental data in any of the following 
three scenarios: i) Effective SUSY models \cite{nath},
ii) Cancellation mechanism \cite{cancell}
and iii) Non-universality of 
trilinear couplings $A_f$ \cite{trilin}.\\
It is well known that the presence of SUSY CP 
violating phases induces mixing 
between the CP--even and 
CP--odd scalars, resulting in the 3 mass eigenstates $H_1^0$,
$H_2^0$ and $H^0_3$ which do not have a definite CP parity.
This mixing affects their phenomenology at present
and future colliders, both in production mechanisms 
and decay partial widths \cite{BR,demir,AA}. 

In this study we will consider various production mechanisms
for neutral Higgs bosons of the MSSM (with and without 
scalar-pseudoscalar mixing) in the context of a NLC with 
$\sqrt s=500$ GeV and $800$ GeV.

{\bf 2.} The study of the various production mechanisms 
of the CP--odd $A^0$ Higgs boson is 
well motivated since the discovery of such a 
particle would signify that the electroweak 
symmetry breaking is introduced by more that one 
Higgs doublet. The CP-odd $A^0$ possesses no
tree--level coupling $A^0$ZZ and $A^0$WW, and so it 
cannot be produced at tree level neither via the
Higgstrahlung process nor via W-W fusion. 
Both of those processes can be generated at one-loop order 
\cite{AAC1,LG,abdes}. 
The one-loop diagrams for $e^+e^-\to ZA^0$ and 
$e^+e^-\to \nu_e \bar{\nu}_e A^0$
can be found respectively in \cite{AAC1,abdes}.
For both processes, the calculation was performed within the 
dimensional regularisation scheme with the help of 
FeynArts and FormCalc \cite{FA}.

For our numerical evaluation, we will take into account the 
following considerations:
The CP conserving MSSM Higgs sector is parametrized by the CP-odd mass 
$M_A$ and $\tan\beta$, taking into account 
radiative corrections to the lightest Higgs boson.
We limit ourselves to the case where $\tan\beta\geq 2.5$.
The chargino/neutralino sector can be parametrized by the 
usual $M_1$, $M_2$ and $\mu$. 
We assume that $M_1\approx M_2/2$ and
$\mu>0$. In our analysis we will take into 
account the following constraints when the SUSY parameters are varied:
i) the extra contribution $\delta\rho$ to the $\rho$ parameter 
\cite{hagiwara} should not exceed the current limits from 
experimental measurements $\delta\rho \la  10^{-3}$, ii) 
$m_{\tilde{t}_1,\tilde{b}_1} > 100$ GeV, $m_{\chi_1^\pm} > 103$ GeV, 
$m_{\chi_1^0} > 50$ GeV and $m_h>110$ GeV.

Following the approach of \cite{GunionKal} we assume a detection 
threshold of 0.1 fb for $e^+e^-\to A^0Z$. This would give 50 events before
experimental cuts for the expected luminosities of 500 fb$^{-1}$. 
In the THDM this criterion would require $\tan\beta\leq 0.3$, even for a 
light pseudoscalar \cite{AAC1}. 

We start by recalling that the THDM contribution
to $e^+e^-\to A^0Z$ \cite{AAC1} in the small $\tan \beta$
regime is enhanced by the top quark contribution, leading to 
cross--sections of order 0.04 fb. In the large 
$\tan \beta$ regime the cross--section is suppressed and does 
not attain observable rates. 
In the MSSM we limit ourself to the case where $\tan\beta\geq 2.5$,
and consequently the THDM contribution is suppressed to the order 
of $\approx 0.002$ fb at $\sqrt s=500$ GeV.
The light SUSY particles (charginos) can slightly enhance the cross section, 
see Fig.~1 (left pannel),
but due to a cancellation between the vertex and box diagrams
the light SUSY enhancement turns out to be not very promising. 
The maximum cross-section for light $M_A$ is about 0.005 fb for 
$\tan\beta=2.5$. As can be seen in Fig.~1, the 
cross-section can also be enhanced by polarizing the electron 
and positron beams. 

\begin{figure}[t!]
\smallskip\smallskip 
\vskip-4.cm
\centerline{{
\epsfxsize2.7 in 
\epsffile{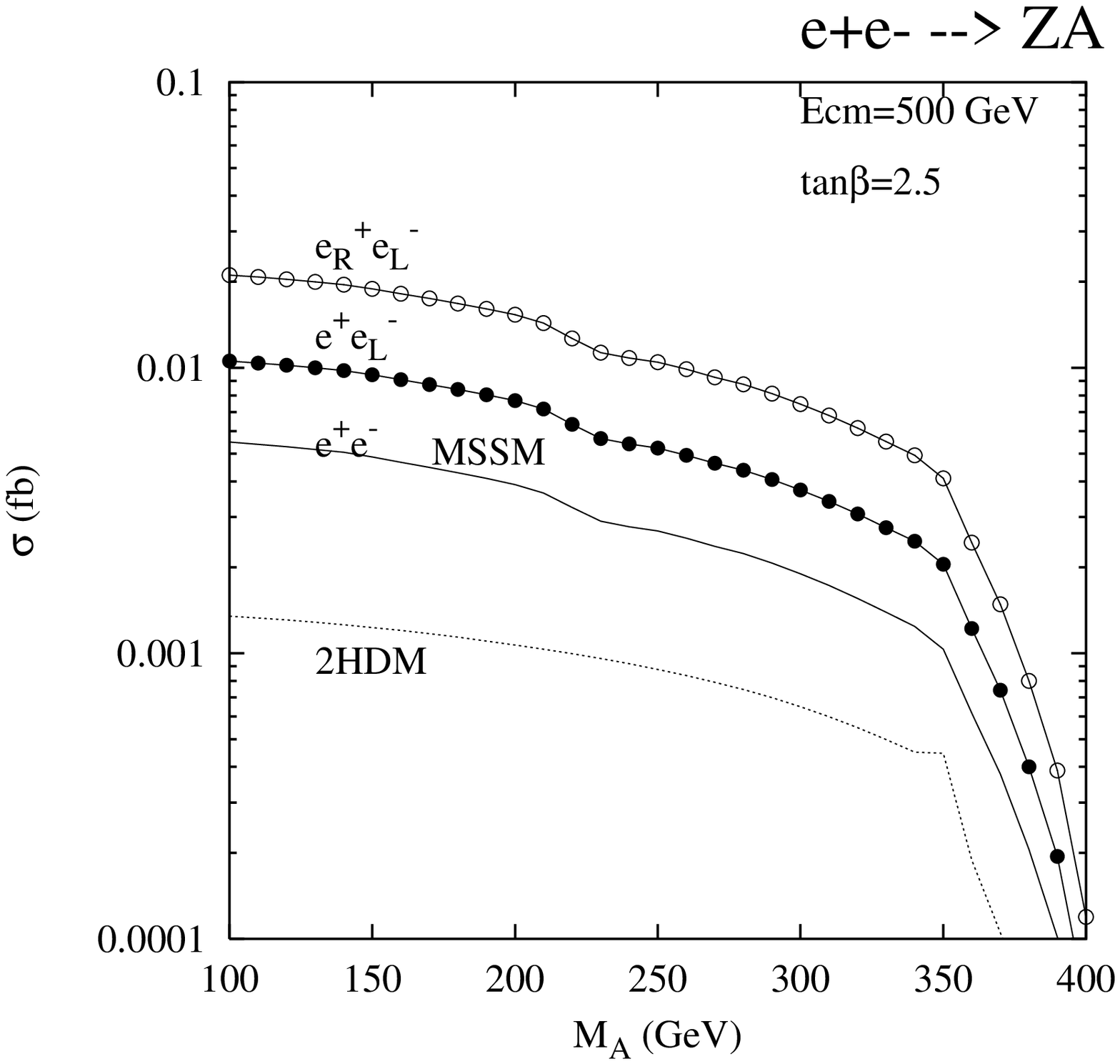}}  \hskip0.4cm
\epsfxsize2.7 in 
\epsffile{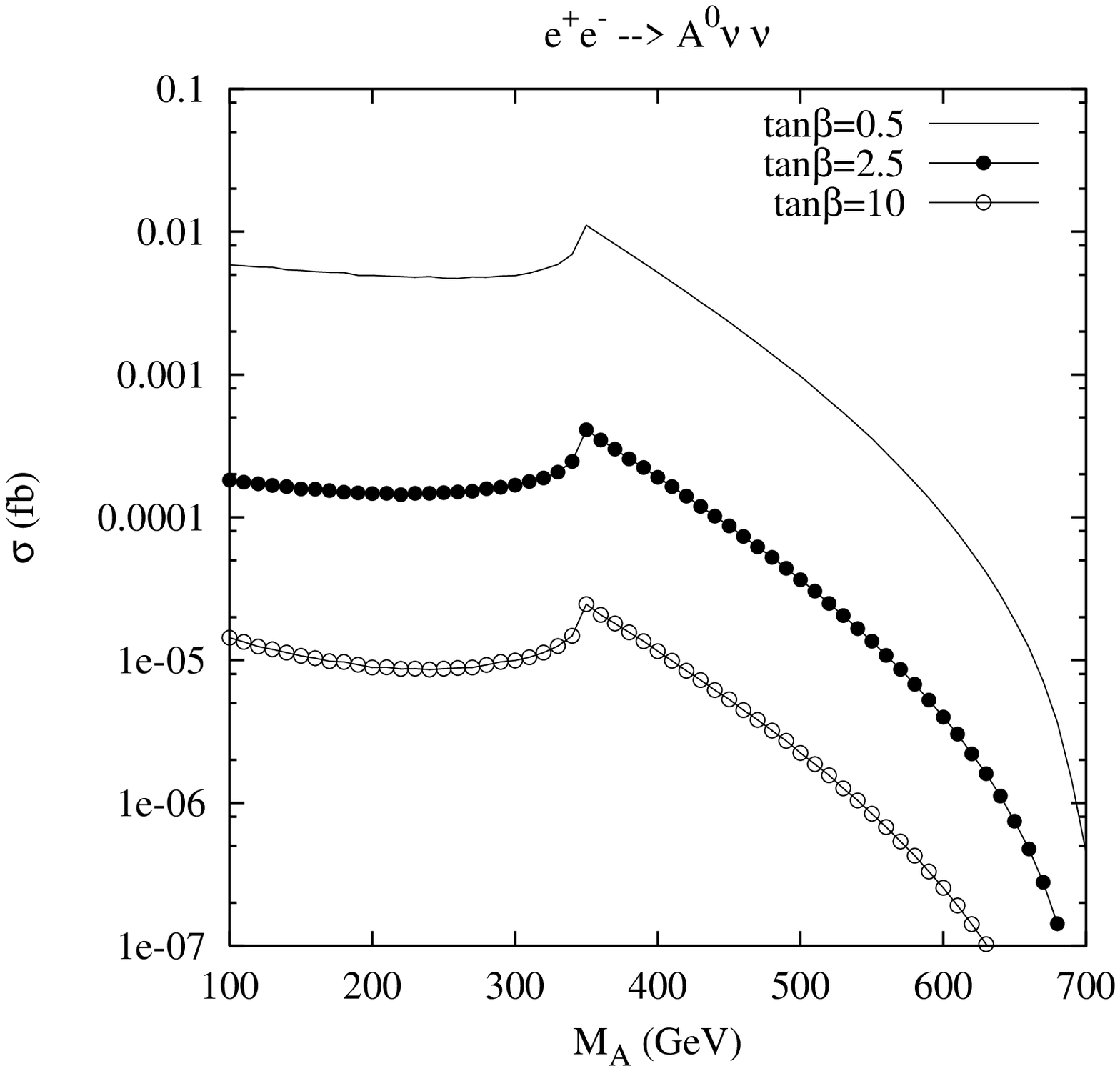}}
\smallskip\smallskip
\caption{Total cross-section for $e^+e^- \to A^0 Z$ 
as function of $M_A$ at $\sqrt{s}=500$ GeV   
(left panel) and for $e^+e^- \to \nu_e \overline \nu_eA^0$
(right panel) at $\sqrt{s}=800$ GeV} \label{fey}
\end{figure}
Let us now discuss the single CP-odd Higgs boson production
$e^+e^- \to \nu_e \bar{\nu}_e A^0$. This one-loop process 
has contributions from: i) one-loop W-W fusion to $A^0$, ii)
the one loop process $e^+e^- \to A^0 Z$ followed by 
Z decay to $\nu_e \bar{\nu}_e $, and iii) box diagrams \cite{abdes}.
In our study we have omitted the five point-functions, 
since these five point-functions do not have any enhancement factor, 
like a top quark or a scalar top quark inside the loop. 
Hence their contribution is expected to be smaller.
Ref. \cite{LG} evaluated the top-bottom contribution to 
$e^+e^- \to \nu_e \bar{\nu}_e A^0$ coming from the one-loop W-W fusion
to $A^0$. 

In Fig.~1 (right panel), we plot the cross-section of 
$e^+e^- \to \nu_e \bar{\nu}_e A^0$
in the 2HDM for 800 GeV center of mass energy. 
For the Higgs couplings and masses we use the MSSM 
values with radiative corrections to the lightest Higgs boson mass.
We found cross-sections of order 0.01 fb (resp 0.0006 fb) for small 
$\tan\beta=0.5$ (resp $\tan\beta=2.5$) and light $M_A<400$ GeV.\\
Numerically, in the MSSM with real parameters, we found that even 
in the optimistic scenario where all
SUSY particles are light (of order 200 GeV), $\tan\beta=2.5$,
large $A_{t,b}$ and large $\tan\beta$,
the cross section does not receive a substantial enhancement.
One concludes that the cross sections for $e^+e^- \to Z A^0$ and
$e^+e^- \to \nu_e \bar{\nu}_e A^0$ are well below $0.1$ fb.

{\bf 3.} We now consider the effect of SUSY CP violating phases on the
Higgs bosons production mechanisms.
We will study the processes $e^+e^-\to Z^*\to H^0_iZ$ 
and $e^+e^-\to H^0_i\nu_e\overline \nu_e$
in the context of a NLC. Both mechanisms are mediated by the tree--level
effective couplings $H^0_iVV$ $(\equiv C_i$), 
but their cross--sections have different 
phase space and $\sqrt s$ dependence. In the CP conserving case one of
these couplings would be zero, corresponding to the absence of 
the coupling $A^0VV$. We will also study the production of 
neutral Higgs pairs, $e^+e^- \to H^0_iH^0_j$ ($i\neq j$). In the CP 
conserving MSSM, only the vertices $Zh^0A^0$ and $ZH^0A^0$ exist at 
tree--level while in the CP violating scenario all three couplings 
$ZH^0_1H^0_2$, $ZH^0_1H^0_3$ and $ZH^0_2H^0_3$ are 
generated at tree--level.
Therefore an observable signal for all $i(=1,2,3)$
in a given mechanism ($e^+e^-\to ZH_i^0$, $e^+e^-\to H_i^0 H_j^0$ or  
$e^+e^-\to H^0_i\nu_e\overline 
\nu_e$), would be a way of probing CP violation in
the Higgs sector. We will calculate the tree--level 
rates of the above mechanisms in the context of the  
MSSM with scalar-pseudoscalar mixing, showing that in the 
most favourable scenarios this way of probing scalar--pseudoscalar 
mixing can be effective if $M_{H_{2,3}}\la 300$ GeV. 

It has been shown \cite{MSSMCP2} that sizeable scalar--pseudoscalar 
mixing is possible for large $|\mu|$,$|A_t| > M_{SUSY}$.
In  \cite{MSSMCP2} the mass matrix ${\cal M}^2_{ij}$ is evaluated 
to one--loop order using effective potential techniques and 
includes large two--loop non--logarithmic corrections 
induced by one-loop threshold effects on the top and bottom 
quark Yukawa couplings. The public code
which we will employ in our numerical analysis  can be found 
in \cite{cph}. 

In the CP conserving MSSM,
$C_1=\sin(\beta-\alpha)$ while one of $C_2$,$C_3$
is identified as $\cos(\beta-\alpha)$, and the other is identically
zero. $C_i\equiv 0$ corresponds to the couplings $A^0ZZ$ 
and $A^0WW$, which will take on a non--zero value at 1-loop. 
Hence to lowest order in the MSSM only $H^0_1$ and {\it one} of $H^0_2$,
$H^0_3$ can be produced in the Higgsstrahlung 
and $WW$ fusion mechanisms, 
$e^+e^-\to Z^*\to H^0_iZ$ and $e^+e^- \to H^0_i\nu_e\overline \nu_e$. 
In the presence of SUSY phases 
all $H^0_i$ may be produced at tree--level via $e^+e^-\to Z^*\to H^0_iZ$,
and an observable signal for all three $H^0_i$ would be evidence for CP 
violation in the Higgs sector.
We stress here that the smallest of 
$\sigma(e^+e^-\to Z^*\to H^0_iZ)$ (resp 
$\sigma(e^+e^-\to W^*W^*\to \nu_e \bar\nu_e H^0_i)$)
 should exceed the 
maximum rate for $e^+e^-\to A^0Z$ (resp
$\sigma(e^+e^- \to \nu_e \bar\nu_e A^0)$)
in the context of the CP conserving MSSM, since the latter
would constitute a ``background'' to any interpretation as 
scalar--pseudoscalar mixing. The previous section showed that
these cross-sections in the CP conserving MSSM
are expected to be comfortably below 0.1 fb.
Note that the process 
$e^+e^- \to H^0_i\nu_e\overline \nu_e$ proceeds via
the same couplings $C_i$, but possesses
a different phase space and $\sqrt s$ dependence. 
This mechanism is competitive with the
Higgsstrahlung process, and becomes the dominant one as
$\sqrt s$ increases. We will also consider the mechanism
$e^+e^-\to H^0_j H^0_k$, which proceeds via the coupling
$ZH_j^0H_k^0$, where $ZZH_i^0=ZH_j^0H_k^0$ for $i\ne j\ne k$. 

In the MSSM (with or without SUSY phases),  
the properties of the lightest eigenstate $H^0_1$ become
very similar to that of the SM Higgs boson in
the decoupling region of $M_{H^\pm}\ge 250$ GeV.  
In this region, $C_1$ is very close to 1, 
and so the sum $C_2^2+C_3^2 $ 
is constrained to be small (a consequence of the sum-rules). 
Therefore we expect that
$M_{H^\pm}\le 250$ GeV will allow larger values for the sum 
$C_2^2 + C^2_3$, and thus observable rates for both $H^0_2$ and
$H^0_3$ in the above mechanisms. 
Distinct signals for all three $H^0_i$ in a given channel 
would be evidence for scalar--pseudoscalar mixing
\footnote{Of course, other models with more than 2 Higgs doublets 
and/or additional singlets could provide multiple signals 
in these channels, e.g. the CP conserving NMSSM.}.
\begin{figure}[t!]
\smallskip\smallskip 
\centerline{{
\epsfxsize3.3 in 
\epsffile{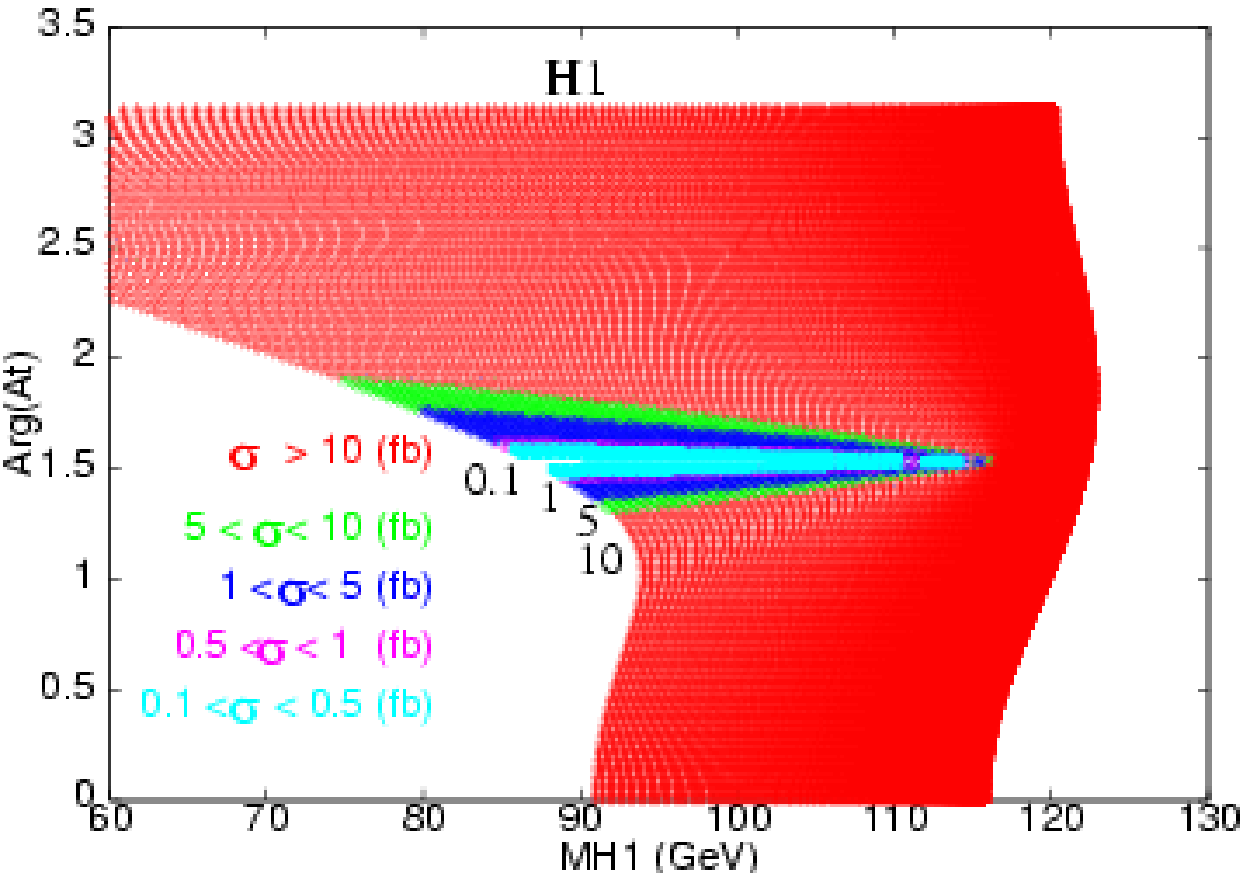}}
\hskip0.4cm
{\epsfxsize3.3 in \epsffile{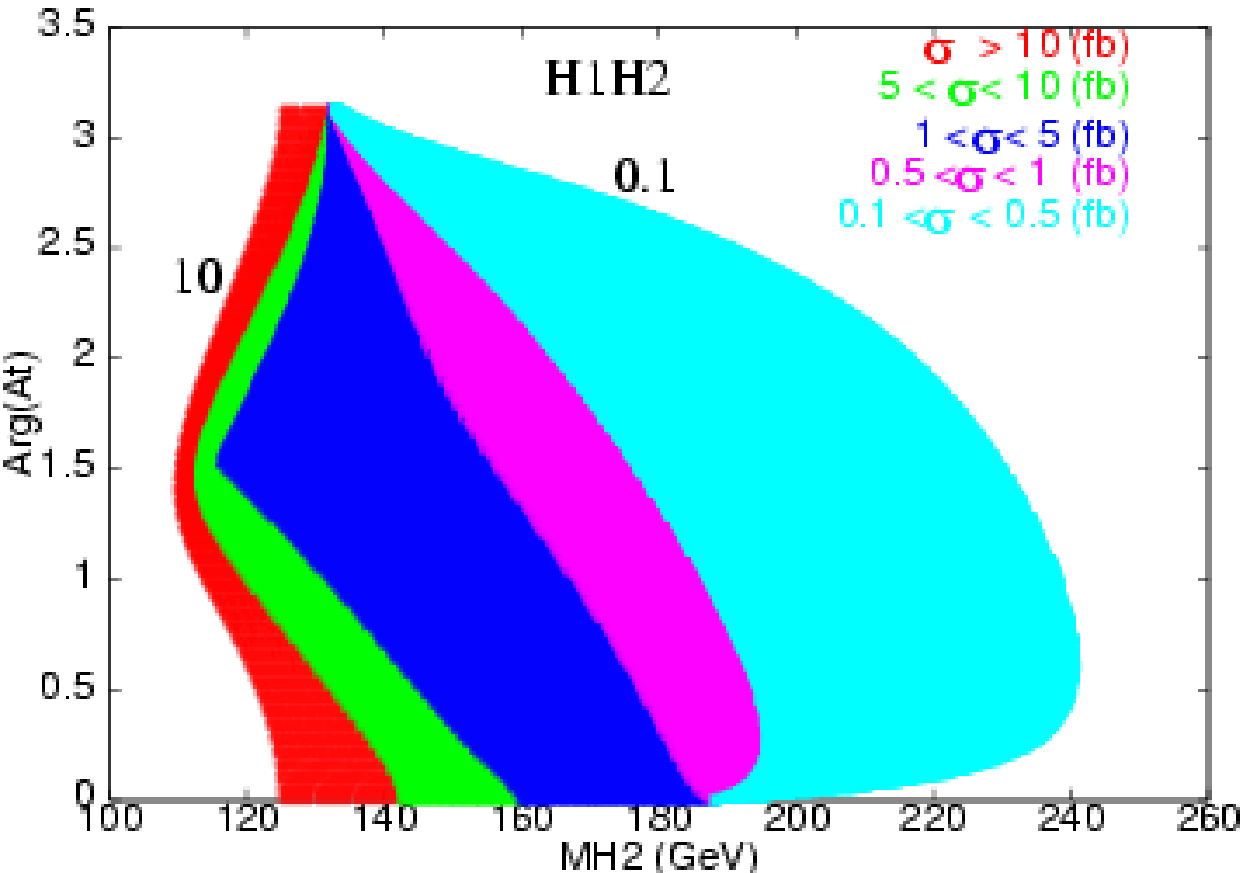}}}
\smallskip\smallskip
\centerline{{
\epsfxsize3.3 in 
\epsffile{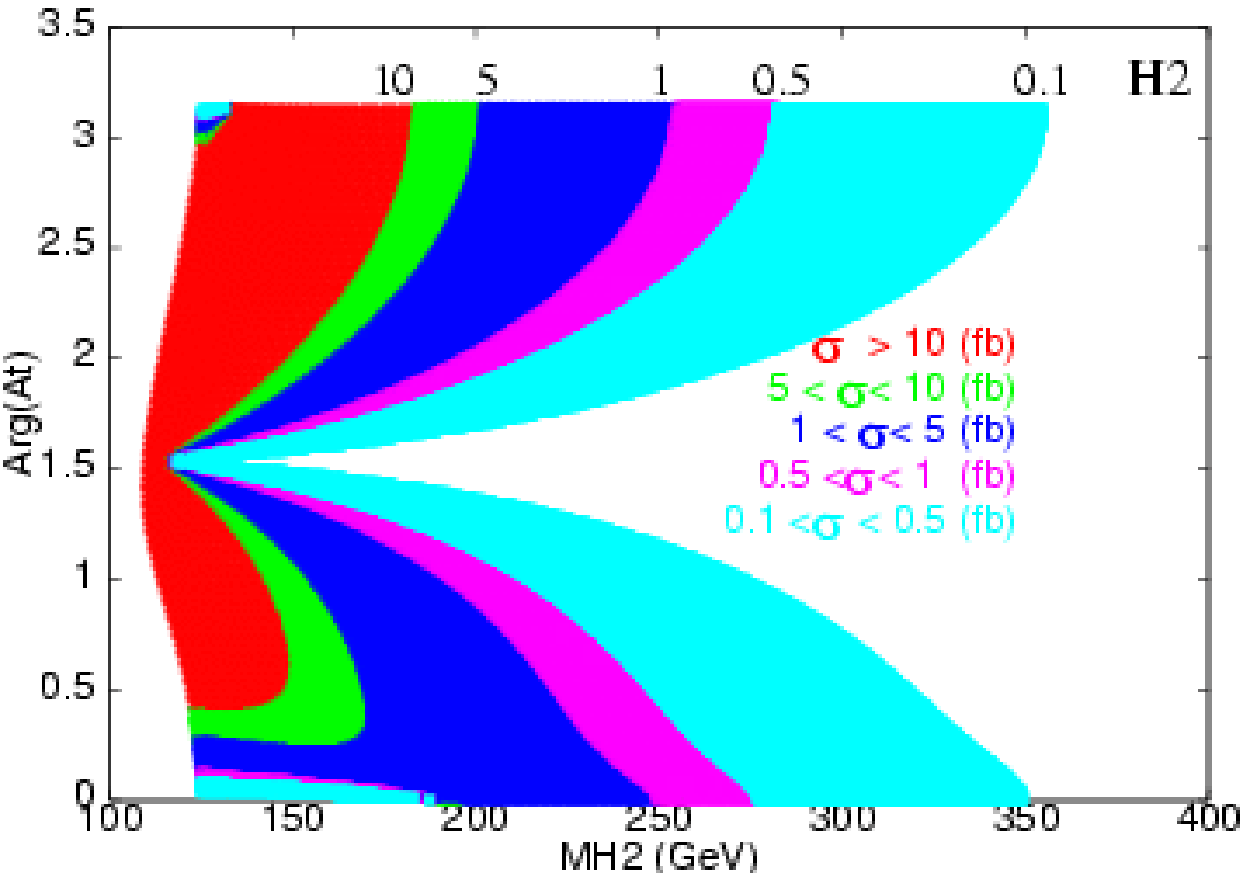}}
\hskip0.4cm
{\epsfxsize3.3 in \epsffile{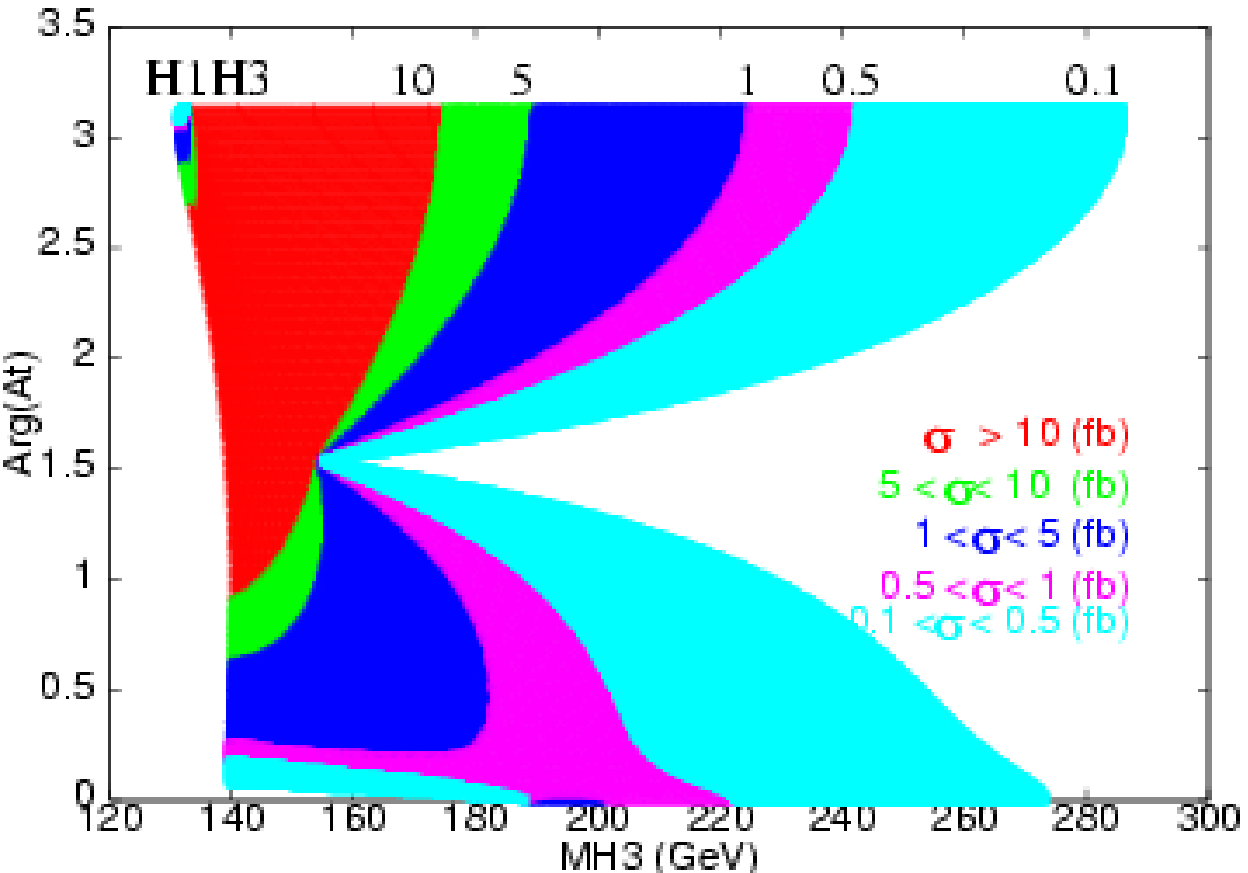 }}}
\smallskip\smallskip
\centerline{{
\epsfxsize3.3 in 
\epsffile{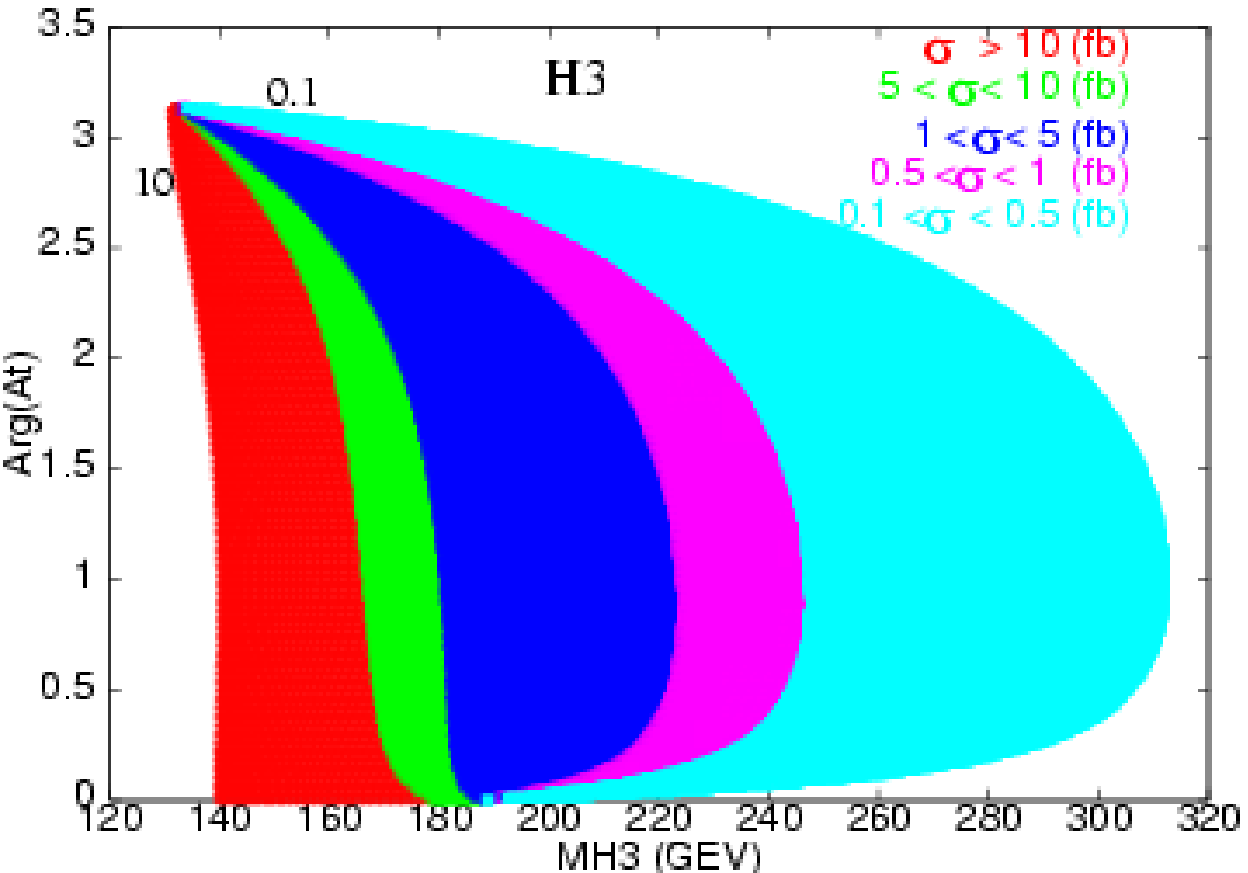}}
\hskip0.4cm
{\epsfxsize3.3 in \epsffile{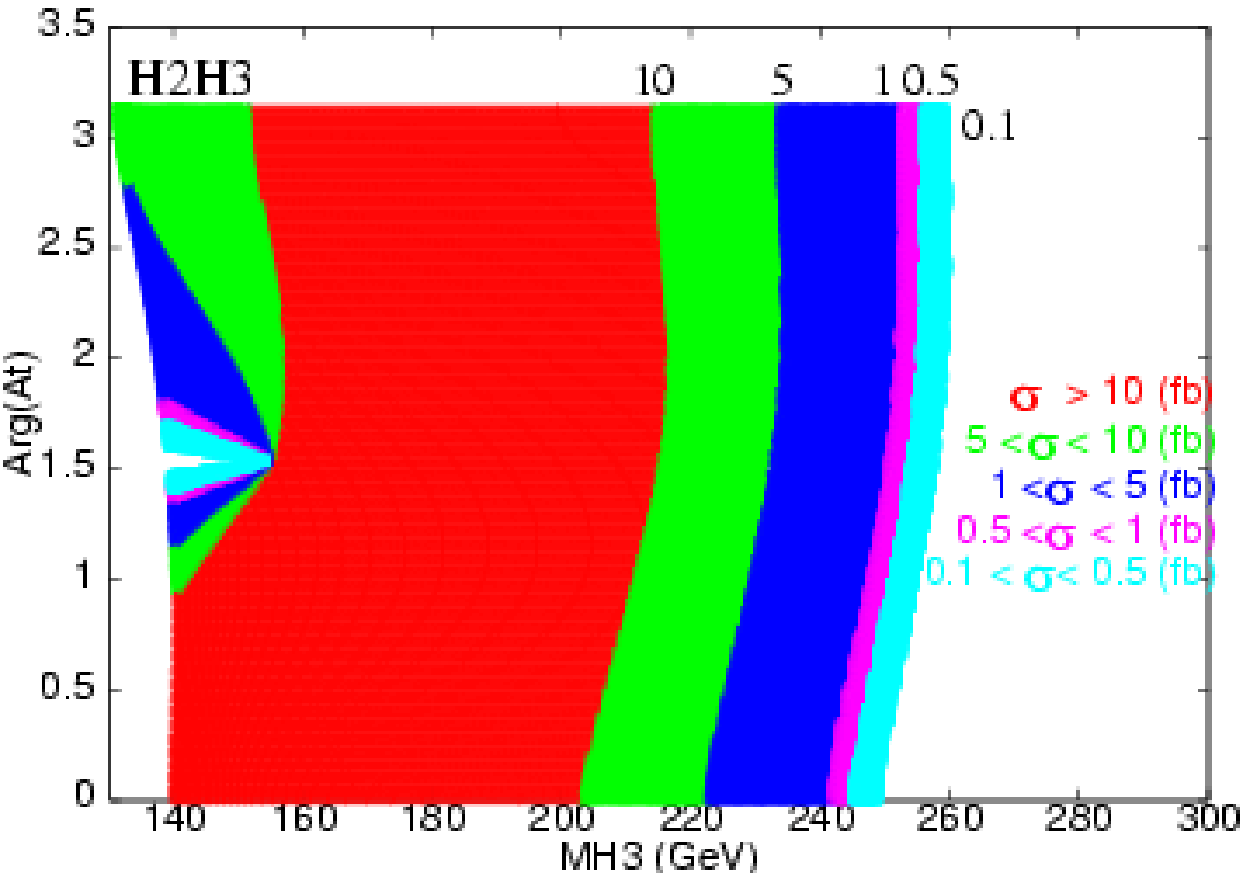 }}}
\smallskip\smallskip
\caption{(left) $\sigma(e^+e^- \to H^0_i \nu\overline{\nu})$ at 
$\sqrt{s}=800$ GeV in ($M_{H_i}$, Arg($A_t$) plane, (right)
$\sigma(e^+e^- \to H^0_iH_j^0)$ at 
$\sqrt{s}=500$ GeV in ($M_{H_i}$, Arg($A_t$) plane}
\label{cros2}
\end{figure}
We now present our numerical results which we will generate with 
the fortran program cph.f \cite{cph}. We note that this program
does not include $\chi^+$--$W$--$H^\pm$ contributions to the 1--loop
neutral Higgs mass matrix, which have been shown to be sizeable in 
some regions of parameter space \cite{nathibrahim}. 

We are concerned with
a NLC collider which has the ability to probe 
$\sigma(e^+e^-\to H^0_iZ,H^0_iH^0_j$,$H^0_i\nu_e\overline \nu)\ge 0.1$ fb,
and so we are also interested
in smaller values for $C^2_2$, $C^2_3$ and larger $M_{H_i}$. 
We shall be presenting results for the production cross--sections
of all the above mechanisms.
Note that the cross--sections for
the processes $e^+e^- \to Zh^0/H^0$, 
$e^+e^- \to Ah^0/H^0$, $e^+e^- \to h^0 H^0$, $e^+e^- \to ZA^0$ and 
$e^+e^- \to \nu_e\overline \nu_e h^0/H^0$ in the CP conserving MSSM
are accurately known \cite{AAC1,hollik,abdel}. Deviations from these rates 
would be evidence for scalar--pseudoscalar mixing.

In our numerical analysis we will choose the CP violating
benchmark scenario (CPX) which maximizes the CP violating effects:
$\widetilde{M}_Q=\widetilde{M}_t=\widetilde{M}_b=M_{SUSY}=1$ TeV, 
$\mu = 4$ TeV, $|A_t|=|A_b|=2$, $|m_{\widetilde{B}}|= 
|m_{\widetilde{W}}|= 0.3$ TeV. 
Note that $\mu$ will be taken real while we allow a CP phase 
in the soft tri-linear parameters $A_t$ and $A_b$. 
The CP phases of $A_t$ and $A_b$ are chosen to be equal. 
In addition, we choose the charged Higgs mass and 
$\tan\beta$ as free parameters.

Our strategy to probe the scalar--pseudoscalar mixing 
requires the identification of the Higgs signals as distinct 
resonances. The inclusion of the phases in $A_t$ and $A_b$ breaks 
the near degeneracy among $M_{H_2}$ and $M_{H_3}$ \cite{MSSMCP2}, 
and gives sufficient splittings to allow identification of 
separate resonances for $H^0_2$ and $H^0_3$. These splittings may
be $>10$ GeV, which is sufficiently large
for a NLC \cite{NLC} to resolve the separate peaks. 
This will lead to three different peaks in the 
Higgsstrahlung and $WW$ fusion processes and motivates us to 
present the individual cross--sections for $e^+e^- \to Z H^0_{i}$ and 
$e^+e^- \to \nu\overline{\nu}_e H^0_{i}$ for $i=1,2,3$.
It has been shown in \cite{MSSMCP2}
that the inclusion of SUSY phases may drastically change the size of 
the couplings $ZZH^0_1$ and $ZH^0_1H^0_2$ for low and 
intermediate $\tan\beta$. 
In such cases the bound on the light Higgs boson obtained at LEPII
may be weakened to $\la 60 $ GeV for large CP violation in the MSSM
Higgs sector. We study the potential of a NLC to discover such a weakly 
coupled Higgs.

In Fig.~2 the left plots depict regions of 
$\sigma(e^+e^- \to \nu_e \bar\nu_e H^0_{i}$) 
in the plane ($M_{H_i}$,{\rm arg}($A_t$))
for $\sqrt s=800$ GeV, $\tan\beta=6$, and $M_{SUSY}=1$ TeV.
In all plots the charged Higgs mass has been varied in increments 
from 140 $\to$ 400 GeV, which determines the values of $M_{H_i}$.
As it can be seen in the plot, $H^0_1$ discovery is 
possible over most of the ($M_{H_1}$,{\rm arg}($A_t$)) 
plane, with small unobservable regions where 
$\sigma(e^+e^- \to Z H^0_{1})< 0.1$ fb which occur
for ${\rm arg}(A_t)\approx 1.5$  and 
$M_{H_1}\la 105$ GeV.

$ZZH_2^0$ is maximized for $M_{H_2}\la 150$ GeV and 
is minimized for ${\rm arg}(A_t) \approx 1.5$ and 
$M_{H_2}\ga 150$ GeV. For $H^0_2$ and $H^0_3$, 
both $\sigma(e^+e^- \to \nu_e\bar\nu_e H^0_{2,3}$) 
can be observable over a wide
region of the plane, even up to relatively large mass values 
e.g. for ${\rm arg}(A_t)=1$,
$\sigma(e^+e^- \to \nu_e \bar\nu_e H^0_{2,3})\ge 0.1$ fb for 
$M_{H_2}\le 275$ and $M_{H_3}\le 310$ GeV. 
Note that the scalar--pseudoscalar 
composition of $H^0_2$ and $H^0_3$ can change with increasing $M_{H_i}$
e.g. for low ${\rm arg}(A_t$), one can see
that $H^0_3$ is dominantly CP-even scalar for low masses, and has a
much larger cross--section than for that for $H^0_2$. 
As $M_{H_{2,3}}$ increases, $H_2^0$ has the larger CP-even scalar
component and may be produced with an observable rate 
for $M_{H_2}\le 350$ GeV.
The cross section for $e^+e^- \to ZH_i$ has a similar shape as the one for 
$e^+e^- \to \nu_e \bar\nu_e H_i$, but with this process one can probe
scalar-pseudoscalar mixing only up to   
$M_{H_2}\le 250$ GeV and $M_{H_3}\le 270$ GeV
both at $\sqrt{s}=500$ and $800$ GeV.

In Fig.~2 (right panels) we show $\sigma(e^+e^- \to H^0_iH^0_j$) in the plane
$(M_{H_i},{\rm arg}(A_t))$ for $\tan\beta=6$, and $M_{SUSY}=1000$ GeV.
Due to its phase space suppression, this mechanism offers smaller 
cross--sections than the Higgsstrahlung and $WW$ fusion processes, 
and consequently it is not as effective at probing the scalar--pseudoscalar mixing.
Using the fact that $C_i^2=C_{jk}^2$ for $i\neq j \neq k$,
the behaviour of pair production $e^+e^- \to H^0_i H^0_j$
can be roughly understood from the rate of the WW fusion process 
$e^+e^- \to \nu \overline{\nu} H^0_k$. As can be seen from the plots,
there are some similarities between $e^+e^-\to H^0_i H^0_j$ and 
$e^+e^-\to \nu \overline{\nu} H^0_k$, for $i\neq j \neq k$.
However, $\sigma(e^+e^- \to H^0_1H^0_{2,3}$) has an observable rate
in the region where the Higgsstrahlung and $WW$ fusion processes
have very suppressed rates.

{\bf 4.} To conclude, we have studied various production mechanisms for neutral
Higgs bosons in the MSSM in the context of a high--energy $e^+e^-$ collider.
We computed the cross--section for the production
mechanisms $e^+e^-\to A^0Z$ and  $e^+e^-\to \nu_e \bar\nu_e A^0$
in the framework of the MSSM. 
Such processes proceed via higher order
diagrams and are strongly model dependent. 
In the MSSM light SUSY particles may give an enhancement 
to the cross--sections but such not sufficient  
to be observable at NLC. 

We then studied the production processes $e^+e^-\to 
H^0_iZ$, $H^0_iH^0_j$ and $H^0_i\nu_e\overline \nu_e$
in the context of the MSSM with SUSY CP violating phases. 
We showed that in a given
channel the cross--section for all $H^0_i$ ($i=1,2,3$) can be observable
at a Next Linear Collider and would provide evidence for 
scalar--pseudoscalar mixing. At $\sqrt s=500$ GeV the coverage of 
$e^+e^-\to H^0_iZ$ and $H^0_i\nu_e\overline \nu_e$ are comparable, with
observable cross--sections for $M_{H_2}\le 250$ GeV and 
$M_{H_3}\le 270$ GeV for the most favourable choice of ${\rm arg}(A_t)$.
At $\sqrt s=800$ GeV, the process  $e^+e^-\to H^0_i\nu_e\overline \nu_e$ 
offers superior coverage,
with a reach up to $M_{H_{2,3}}\le 300$ GeV in the most favourable cases.
The scalar--pseudoscalar mixing causes a mass splitting 
between $H^0_2$ and $H^0_3$ which should be sufficient for separate peaks
to be resolved at a NLC. The mechanism
$e^+e^-\to H^0_1H^0_{2,3}$, while less effective for probing 
scalar--pseudoscalar mixing,
can comfortably detect $H^0_1$ in the region of suppressed
coupling $VVH^0_1$.


\begin{thebibliography}{99}
\bibitem{MSSMCP2} A.~Pilaftsis,
Phys.\ Lett.\ B {\bf 435} (1998) 88.
D.~A.~Demir,
Phys.\ Rev.\ D {\bf 60}, 055006 (1999);
A.~Pilaftsis et al,
Nucl.\ Phys.\ B {\bf 553}, 3 (1999). 
S.~Y.~Choi et al,
Phys. Lett. {\bf B481} (2000) 57;
M.~Carena et al,
Nucl.\ Phys.\ B {\bf 586} (2000) 92; 
S.~Heinemeyer,
Eur.\ Phys.\ J.\ C {\bf 22}, 521 (2001)
T.~Ibrahim et al,
arXiv:hep-ph/0204092.

\bibitem{baek} 
S.~w.~Baek et al,
Phys.\ Rev.\ Lett.\  {\bf 83}, 488 (1999);
T.~Goto et al,
Phys.\ Lett.\ B {\bf 460}, 333 (1999);
C.~K.~Chua et al,
Phys.\ Rev.\ D {\bf 60}, 014003 (1999);
A.~G.~Akeroyd et al,
Phys.\ Lett.\ B {\bf 507}, 252 (2001)


\bibitem{nath} P.~Nath,
Phys.\ Rev.\ Lett.\  {\bf 66} (1991) 2565; 
Y.~Kizukuri et al,
Phys.\ Rev.\ D {\bf 46}, 3025 (1992).

\bibitem{cancell} T.~Ibrahim et al,
Phys.\ Lett.\ B {\bf 418} (1998) 98;
Phys.\ Rev.\ D {\bf 57} (1998) 478
[Erratum-ibid.\ D {\bf 58} (1998) 019901;
\ D {\bf 60} (1999) 079903;\ D {\bf 60} (1999) 119901]; 
M.~Brhlik et al,
Phys.\ Rev.\ D {\bf 59} (1999) 115004.

\bibitem{trilin}
S.A. Abel and J.M. Frere,
Phys.\ Rev.\ D {\bf 55} (1997) 1623.

\bibitem{BR} A.~Dedes et al,
Nucl.\ Phys.\ B {\bf 576} (2000) 29;
S.~Y.~Choi et al,
Phys.\ Rev.\ D {\bf 61} (2000) 115002;
J.~L.~Kneur et al,
Phys.\ Rev.\ D {\bf 61}, 095003 (2000);
S.~Y.~Choi et al,
Phys.\ Rev.\ D {\bf 64}, 032004 (2001);
A.~Arhrib et al,
Phys.\ Lett.\ B {\bf 537}, 217 (2002);


\bibitem{demir}
D.A.~Demir,
Phys.\ Lett.\ B {\bf 465}, 177 (1999);
S.W.Ham et al,
J.\ Phys.\ G {\bf 27} (2001) 1.

\bibitem{AA} A.~G.~Akeroyd and A.~Arhrib,
Phys.\ Rev.\ D {\bf 64}, 095018 (2001)

\bibitem{AAC1} 
A.~G.~Akeroyd et al,
Phys.\ Rev.\ D {\bf 64}, 075007 (2001), E-ibid.\ D {\bf 65}, 099903 (2002)
A.~G.~Akeroyd et al,
Mod.\ Phys.\ Lett.\ A {\bf 14}, 2093 (1999), E-ibid.\ A {\bf 17}, 373 (2002)

\bibitem{LG} 
T.~Farris, J.~F.~Gunion and H.~E.~Logan,
arXiv:hep-ph/0202087.

\bibitem{abdes}
A. Arhrib, hep-ph/0207330

\bibitem{FA} J.~Kublbeck et al,
Comput.\ Phys.\ Commun.\  {\bf 60}, 165 (1990);
T.~Hahn,
Comput.\ Phys.\ Commun.\  {\bf 140}, 418 (2001);
T.~Hahn et al,
Comput.\ Phys.\ Commun.\  {\bf 118}, 153 (1999);

\bibitem{hagiwara}
M.~Drees and K.~Hagiwara,
Phys.\ Rev.\ D {\bf 42}, 1709 (1990).

\bibitem{GunionKal} 
B. Grzadkowski et al, Phys. Rev. Lett. {\bf 79} (1997) 982;
B. Grzadkowski et al, Phys. Rev. {\bf D60}
(1999) 075011; 

\bibitem{cph} http://home/cern.ch/p/pilaftsi/www/.

\bibitem{nathibrahim} T.~Ibrahim et al,
Phys.\ Rev.\ D {\bf 63} (2001) 035009; {\sl ibid}
arXiv:hep-ph/0204092.

\bibitem{hollik} 
S. Heinemeyer et al hep-ph/0102117;
 S. Heinemeyer et al,
Eur.Phys.J.C {\bf 19},535 (2001)

\bibitem{abdel} 
A.~Djouadi, V.~Driesen, W.~Hollik and J.~Rosiek,
Nucl.\ Phys.\ B {\bf 491}, 68 (1997)


\bibitem{NLC} E.~Accomando {\it et al.}  
Phys.\ Rept.\  {\bf 299}, 1 (1998).


\end{thebibliography}
\end{document}